\documentclass[sigconf,natbib=true,anonymous=false]{acmart}

\setcopyright{none}
\renewcommand\footnotetextcopyrightpermission[1]{}
\definecolor{ao(english)}{rgb}{0.0, 0.5, 0.0}

\usepackage[inline]{enumitem}

\usepackage{glossaries-extra}
\setabbreviationstyle[no-long]{short}  

\glsdisablehyper

\usepackage{physics}
\usepackage{tikz}
\usepackage{mathdots}
\usepackage{cancel}
\usepackage{color}
\usepackage{siunitx}
\usepackage{array}
\usepackage{multirow}
\usepackage{gensymb}
\usepackage{tabularx}
\usepackage{extarrows}
\usepackage{booktabs}
\usetikzlibrary{fadings}
\usetikzlibrary{patterns}
\usetikzlibrary{shadows.blur}
\usetikzlibrary{shapes}

\newcommand{\para}[1]{\smallskip\noindent{\bf{#1}}}

\usepackage{array}
\newcolumntype{R}{>{\setbox0=\hbox\bgroup}c<{\egroup}@{}}
\usepackage{numprint}
\usepackage{pifont} 
\usepackage{float}
\usepackage[section]{placeins}
\usepackage{stfloats}

\usepackage[ruled,vlined, linesnumbered]{algorithm2e}
\usepackage{amsmath}

\usepackage{caption}
\usepackage{subcaption}

\setlist[itemize]{leftmargin=*,topsep=0pt}
\newacronym{kg}{KG}{Knowledge Graph}
\newacronym{kge}{KGE}{Knowledge Graph Embedding}
\newacronym{ckg}{CKG}{Collaborative Knowledge Graph}

\newacronym{gnn}{GNN}{Graph Neural Network}
\newacronym{rs}{RS}{Recommender System}
\newacronym{cf}{CF}{Collaborate Filtering}
\newacronym{rnn}{RNN}{Recurrent Neural Network}
\newacronym{nlp}{NLP}{Natural Language processing}
\newacronym{bpr}{BPR}{Bayesian Personalized Ranking}
\newacronym{cnn}{CNN}{Convolutional Neural Network}
\newacronym{gcn}{GCN}{Graph Convolutional Network}
\newacronym{asha}{ASHA}{Asynchronous Successive Halving}

\newacronym{toppop}{TopPop}{Top Popular}
\newacronym{ppr}{PPR}{Personalized PageRank}
\newacronym{bprmf}{BPR-MF}{Bayesian Personalized Ranking for Matrix Factorization}
\newacronym{kgat}{KGAT}{Knowledge Graph Attention Network}
\newacronym{ae}{AE}{AutoEncoder}
\newacronym[category=no-long]{own}{SimpleRec}{Simple and strong inductive baseline for Recommendation}
\newacronym{mpnn}{MPNN}{Message Passing Neural Network}
\newacronym{ngcf}{NGCF}{Neural Graph Collaborative Filtering}
\newacronym{lgcn}{LGCN}{LightGCN}

\newacronym{mr}{MR}{MindReader}
\newacronym{ml}{ML-20m}{MovieLens-20m}
\newacronym{mls}{ML-S}{MovieLens Subsampled}
\newacronym{ab}{AB}{Amazon-Book (2014)}
\newacronym{abs}{AB-S}{Amazon-Book Subsampled}
\newacronym{fm}{FM}{LFM-1b}
\newacronym{syn}{SYN}{Synthetic-1m}

\newacronym[longplural={Recommendable Entities}]{re}{RE}{Recommendable Entity}
\newacronym[longplural={Descriptive Entities}]{de}{DE}{Descriptive Entity}
\newacronym{sota}{SotA}{State-of-the-Art}

\begin{document}

\title{Simple and Powerful Architecture for Inductive Recommendation Using Knowledge Graph Convolutions}

\author{Theis E. Jendal, Matteo Lissandrini, Peter Dolog, Katja Hose}
\email{{tjendal,matteo,dolog,khose}@cs.aau.dk}
\affiliation{%
  \institution{Department of Computer Science, Aalborg University}
  \country{Denmark}
}

\begin{abstract}
%
Using graph models with relational information in recommender systems has shown promising results.
Yet, most methods are \emph{transductive}, i.e., they are based on dimensionality reduction architectures.
Hence, they require heavy retraining every time new items or users are added.
Conversely, \emph{inductive methods} promise to solve these issues.
Nonetheless, all inductive methods rely only on interactions, making recommendations for users with few interactions sub-optimal and even impossible for new items. 
Therefore, we focus on inductive methods able to also exploit knowledge graphs (KGs).
In this work, we propose \gls{own}, a strong baseline that uses a graph neural network and a KG to provide better recommendations than related inductive methods for new users and items. 
We show that it is unnecessary to create complex model architectures for user representations, but it is enough to allow users to be represented by the few ratings they provide and the indirect connections among them without any user metadata. 
As a result, we re-evaluate state-of-the-art methods, identify better evaluation protocols, highlight unwarranted conclusions from previous proposals, and showcase a novel, stronger baseline for this task.
%
%

%
\end{abstract}




\maketitle
\pagestyle{plain}

\newcommand{\verts}{\mathcal{V}}  
\newcommand{\edges}{\mathcal{E}}  
\newcommand{\users}{\mathcal{U}}  
\newcommand{\warmusers}{\mathcal{U}_w}  
\newcommand{\coldusers}{\mathcal{U}_c}  
\newcommand{\feedback}{\mathcal{C}}  

\newcommand{\entities}{\mathcal{N}}  
\newcommand{\warmrecs}{\mathcal{I}_w}  
\newcommand{\coldrecs}{\mathcal{I}_c}  
\newcommand{\items}{\mathcal{I}}  
\newcommand{\recs}{\mathcal{I}}  
\newcommand{\descs}{\mathcal{E}_{desc}}  
\newcommand{\labels}{\mathcal{L}}  
\newcommand{\graph}{\mathcal{G}}  
\newcommand{\relations}{\mathcal{R}}  
\renewcommand{\real}{\mathbb{R}}  

\newcommand{\interactions}[1]{\mathbf{I}_{#1}}

\newcommand{\user}{u}
\newcommand{\userother}{v}

\newcommand{\rec}{i}
\newcommand{\otherrec}{j}

\newcommand{\entity}{{e}}
\newcommand{\otherentity}{{e_1}}

\newcommand{\head}{h}
\newcommand{\relation}{r}
\newcommand{\tail}{t}

\newcommand{\ratingfunc}{R}
\newcommand{\globalpr}{S}
\newcommand{\loss}{\mathbf{L}}
\newcommand{\Q}[1]{Q^{#1}}
\newcommand{\relmapping}{\phi}
\newcommand{\mse}{\text{MSE}}
\newcommand{\func}{\mathcal{F}}
\newcommand{\feature}{\mathcal{X}}

\newcommand{\cvector}[1]{\mathbf{#1}}
\newcommand{\cmatrix}[1]{\mathbf{#1}}
\newcommand{\preference}[1]{\leqslant_{#1}}
\newcommand{\predpref}[1]{\:\widehat{\preference{#1}}\:}
\newcommand{\topn}{top-$n$}
\newcommand{\pipe}{\bigm|}
\newcommand{\iembedding}{\cmatrix{X}}
\newcommand{\hidden}[1]{\cmatrix{\entity}_{#1}}
\newcommand{\lae}{K} 

\RestoreAcronyms
\section{Introduction}
\label{sec:introduction}

Many \glspl{rs} 
identify user preference patterns assuming that users with similar past behavior have similar preferences, e.g., users that 
watch the same movies are likely to do so also in the future; this approach is commonly referred to as \gls{cf}~\cite{he2017neuralncf, cremonesi2010topnperformance,  he2020lightgcn, rendle2012bpr, wang2019neuralngcf}. 
Usually, this translates to embedding users and items into a low-dimensional space, where the representation of a user is similar to the representations of the items that are more likely to be relevant to them.
Yet, \emph{many existing methods only work in a transductive setting}, where it is assumed that all users and items have been seen during training~\cite{hamilton2017inductive, zhang2021inductiveicp}.
In contrast, \emph{in an inductive setting, users and items exist that are not in the training set}.
The method should be able to provide predictions on these users and items when information about them is acquired.
This is typical in real-world online recommendation scenarios where users and items are continuously added.
An inductive method therefore \emph{does not require retraining each time a new user, item, or rating is added} to the system like transductive methods do; instead, it can immediately reason about the newly added information and update its predictions.

\begin{figure}[!tb]
    \centering
    \includegraphics[clip, trim=100 50 900 0, width=0.87\linewidth]{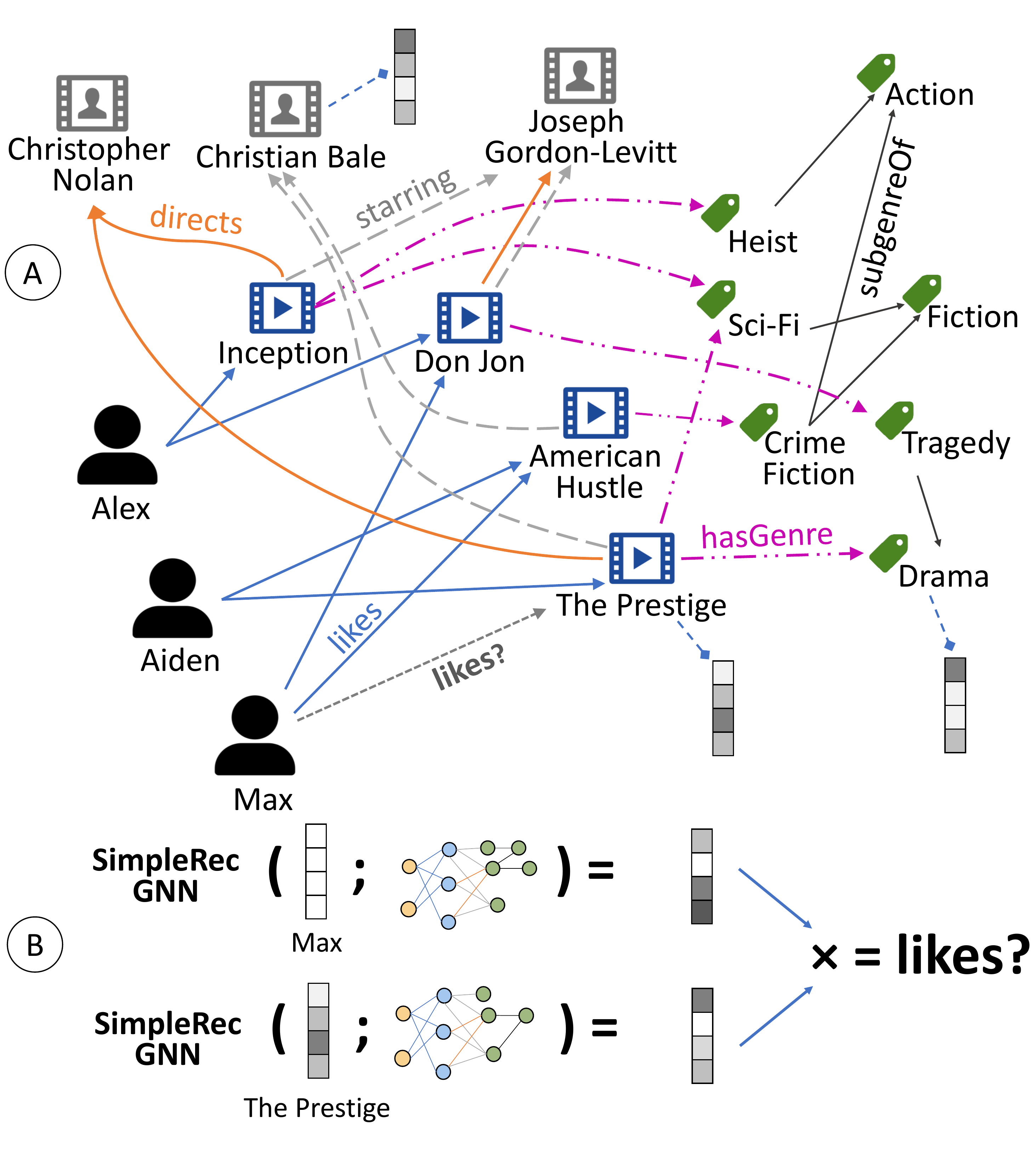}
    \vspace*{-16pt}
    \caption{Item recommendation over a CKG:\,Part\,A shows a CKG in the movie domain, with users, movies, and connected entities; Part\,B illustrates the recommendation task.}
    \label{fig:example}
    \vspace*{-18pt}
\end{figure}

However, current inductive methods~\cite{ying2018graphpinsage,zhang2019inductiveigmc,wu2021towardsidcf,wang2021priviledgepgd,zhang2022geometricgimc,zhang2021inductiveicp} only use interaction data, such as ratings, making them unable to handle situations where this type of data is sparse, e.g., long-tail users and items.
Instead, including \gls{kg} information would allow to handle long-tail users and items.
Hence, we propose \gls{own}, a new simple-yet-powerful architecture that uses \gls{kg} for inductive predictions in a scalable way.  
Our baseline respects the following 4 important tenets: 
\begin{enumerate*}[label=(\roman*)]
    \item it does not require any user metadata at all (e.g.,  personal attributes),
    \item it reduces the effect of the popularity bias by including both user-item information and item-entities connections\footnote{\label{fn:app_abl}The results in \autoref{app:abl} illustrates this.},
    \item it ensures scalability by avoiding the creation of any subgraph for each user-item pair and avoiding other complex sampling methods, and
    \item it exploits both relational information\textsuperscript{\ref{fn:app_abl}} as well as any additional contextual information, e.g., textual attributes.
\end{enumerate*}
As a result \gls{own} sets itself apart from trivial extensions of the GraphSAGE architecture as well as any other existing inductive method~\cite{ying2018graphpinsage,wang2019kgat, wang2018ripplenet, yang2020hagerec, palumbo2020entity2rec, wang2019explainableknowledgegraphkprn,zhang2019inductiveigmc,wu2021towardsidcf,wang2021priviledgepgd,zhang2022geometricgimc,zhang2021inductiveicp}.
We adopt this new strong baseline to re-evaluate state-of-the-art inductive recommendation methods.
In doing so, we also put under scrutiny existing experimental protocols performed in previous studies.
As a result, we identify also inconsistencies in previous evaluation protocols.
Thus, \emph{our experiments show both a better way to evaluate Inductive Recommendation methods using KGs as well as an important scientific direction that has been evidently overlooked}.

In the following, after providing the definition of recommendation over collaborative KGs, we illustrate issues with state-of-the-art methods for inductive recommendations, then we describe our proposed architecture, which we publish open-source along with the evaluation suite and re-implemented state-of-the-art methods at~\url{https://anonymous.4open.science/r/SimpleRec-DB4F}. 
Finally, we present our evaluation methodology, overcoming existing issues, as well as the promising performances of our new proposed baseline. 
\section{Collaborative Knowledge Graphs} 
\label{sec:problem_formulation}
A \gls{kg} is a heterogeneous graph ${\graph}:{\langle}{\verts},{\edges}{\rangle}$ representing entities as vertices (also nodes) $\verts$ and the semantic relations connecting them as labeled edges $\edges:\verts{\times}\relations{\times}\verts$, given the relation types $\relations$. 
In \autoref{fig:example}, nodes represent recommendable items (${\items}{\subset}{\verts}$), i.e., movies, and their connected entities (${\entities}{\subset}{\verts}$), e.g.,  actors, directors, and a taxonomy of genres, s.t. $\entities {\cap} \items {=} \emptyset$.
Edges represent how nodes are connected, e.g., the entity ``Inception'' is connected to the genre ``Heist'' through the relation ``has genre''.
Furthermore, we adopt the concept of a \gls{ckg}~\cite{wang2019kgat}, i.e., a knowledge graph augmented with users (${\users}{\subset}{\verts}$) and user interactions, e.g., (\textsf{Max}, \textsf{likes}, \textsf{Don Jon}).
We note that linking nodes in a KG also gives access to additional information that can be attached to nodes, e.g., encyclopedic texts about entities or product descriptions for items.
Conversely, we explicitly avoid to assume any additional personal information about users, e.g., age or gender, as this information is usually not available and its storage constitutes more often a liability because of privacy concerns~\cite{wang2018toward}.

In practice, we model our task as a \emph{top-k recommendation problem}. 
We aim at learning a model that ranks items according to a user's preferences estimated as $\hat{y}$; i.e., $\hat{y}_{\user\rec}{>}\hat{y}_{\user\otherrec}$, if user $\user$ prefers item $\rec$ over $\otherrec$.
Furthermore, in line with previous literature~\cite{wu2021towardsidcf,zhang2019inductiveigmc,lee2019melu}, we define two types of users: warm-start users $\warmusers$ where some interactions are known at training time and cold-start users $\coldusers$, for which no interaction is known at training time, \emph{but some become known at inference time}.
Thus, we only assume that entities in the KG itself remain unchanged, and we evaluate methods able to provide recommendation for the cold-start users.

\vspace*{-7pt}
\section{Related Work}
\label{sec:relatedWork}

\setlength{\textfloatsep}{0.6cm}
\definecolor{myGreen}{RGB}{77, 175, 74}
\definecolor{amber}{rgb}{1.0, 0.75, 0.0}

\newcommand{\cding}[2]{\textcolor{#1}{\ding{#2}}}
\newcommand{\yes}{\cding{myGreen}{52}}
\newcommand{\no}{\cding{red}{55}}
\newcommand{\maybe}{\color{amber}(\ding{52})}

\begin{table}[!tb]
\caption{Related methods,  whether they use User Metadata, whether they handle Relational information (i.e., KG), the Task they support among (C) Node Classification, (R) Ranking, and (P) Rating Prediction, and whether the method constructs a Subgraph from user-item pairs.}
\vspace{-12pt}
\resizebox{\linewidth}{!}{%
\begin{tabular}{l|cccccRRc}
\hline
& \multicolumn{2}{c}{\textbf{Inductive}} & \textbf{User} &&& \multicolumn{2}{R}{\textbf{Embedding}} &  \\
\textbf{Model} & \textbf{User} & \textbf{Item} & \textbf{Metadata } & \textbf{Relational} & \textbf{Task} & \multicolumn{2}{R}{$\users~/~\descs$} & \textbf{Subgraph} \\ \hline
NGCF~\cite{wang2019neuralngcf} & \no & \no & \no & \no & R & \yes & \yes & \no \\
KGAT~\cite{wang2019kgat} & \no & \no & \no & \yes & R & \yes & \yes & \no \\
KPRN~\cite{wang2019explainableknowledgegraphkprn} & \no & \no & \no & \yes & R & \yes & \yes & \no \\
MeLU~\cite{lee2019melu} & \no & \no & \yes & \no & R & \yes & \yes & \no \\
LGCN~\cite{he2020lightgcn} & \no & \no & \no & \no & R & \yes & \yes & \no \\ \hline
GraphSAGE~\cite{hamilton2017inductive} & \maybe & \yes & \no & \no & C & \no & \yes & \no \\
PinSAGE~\cite{ying2018graphpinsage} & \maybe & \yes & \no & \no & R & \no & \no & \no \\
IGMC~\cite{zhang2019inductiveigmc} & \yes & \yes & \no & \no & P & \no & \no & \yes \\
IDCF~\cite{wu2021towardsidcf} & \yes & \no & \no & \no & P & \no & \no & \no \\
PGD~\cite{wang2021priviledgepgd} & \yes & \yes & \yes & \no & R & \no & \no & \no \\
ICP~\cite{zhang2021inductiveicp} & \no & \yes & \no & \no & R & \no & \no & \no \\
GIMC~\cite{zhang2022geometricgimc} & \yes & \yes & \no & \no & P & \no & \no & \yes \\
\hline
\textbf{SimpleRec} & \yes & \yes & \no & \yes & R & \yes & \yes & \no \\
\hline
\end{tabular}
}
\label{table:methods}
\vspace*{-12pt}
\end{table}

Most existing recommendation methods are not able to represent nodes that were not present during training and are also limited in handling sparse datasets.
First, most models are transductive, i.e., they embed  each node in the graph within the same low-dimensionality space~\cite{wang2019kgat,wang2019explainableknowledgegraphkprn}. 
Second, most of them utilize only bipartite graphs of user interactions with items~\cite{he2020lightgcn,wang2019neuralngcf,ying2018graphpinsage}.
Instead, inductive learning models generate predictions for unseen nodes by directly reasoning over the features that describe them.
Here, we provide an overview of inductive methods, detailing their limitations as compared to our proposal (as summarized in Table~\ref{table:methods}).

GraphSAGE~\cite{hamilton2017inductive} was the first inductive \gls{gnn} able to efficiently generate embeddings for unseen nodes by leveraging node features, e.g., textual attributes.
It was later modified for the recommendation task in PinSAGE~\cite{ying2018graphpinsage}, also using MapReduce to scale the computation.
Unfortunately, PinSAGE was designed for item-item recommendations, i.e., assuming we have boards of items and want to add a new item to a board. 
The method therefore does not explicitly handle user-item ratings but assumes that items closer in an embedding space will be rated similarly.

Other methods have been proposed for inductive matrix completion~\cite{xu2013speedupimc,jain2013provableimc}, most recently IGMC~\cite{zhang2019inductiveigmc} and GIMC~\cite{zhang2022geometricgimc}, which extract subgraphs around each user-item pair to obtain the necessary representations, passing such sub-graphs through multiple layers of a \gls{gnn}.
These approaches are designed for the single rating-prediction task and not for the ranking task.
\emph{Generating these sub-graphs is prohibitively space- and time-consuming}.
Thus, they cannot efficiently produce user-item rankings, since a sub-graph is generated for \emph{all} user-item pairs\footnote{Similar to previous work~\cite{wu2021towardsidcf} despite our best efforts, we were unable to train and test GIMC and IGMC in the ranking settings due to their running time requirements.}.
Furthermore, these methods do not use KG information; thus, they cannot provide predictions for new items with no interactions. 
Therefore, instead of constructing subgraphs, we employ subsampling of neighboring nodes to obtain a scalable prediction mechanism~\cite{ying2018graphpinsage,hamilton2017inductive} and we use \glspl{kg} to gain information also about items with few user interactions.

Some methods, e.g., PDG~\cite{wang2021priviledgepgd} and MeLU~\cite{lee2019melu}, exploit instead user metadata, e.g., gender and age information.
Yet, this information is rarely available, making it impossible to use these methods in almost all cases\footnote{We excluded MeLU as our tests without user metadata led to very poor results.}. 
Moreover, MeLU~\cite{lee2019melu} retrains the model for each user, which is not scalable.
Hence, in our method, we assume no user metadata, learning instead \textit{how} to aggregate information.

Related works therefore either:
\begin{enumerate*}[label=(\roman*)]
    \item create subgraphs, which do not scale in the ranking task;
    \item use personal user data, which is almost never available; or
    \item predict a single user rating, which under-performs in the ranking task, even compared to non-personalized methods~\cite{cremonesi2010topnperformance}.
\end{enumerate*}
Hence, (\autoref{table:methods}) only GraphSAGE~\cite{hamilton2017inductive}, PinSAGE~\cite{ying2018graphpinsage} and IDCF~\cite{wu2021towardsidcf} are feasible methods to recommend for users that have not been seen during training, but they do not exploit KGs.
Thus, \emph{we are the first to propose a scalable inductive method for user-personalized recommendation} that learns to extract knowledge from a KG while not requiring any user metadata.

\vspace*{-7pt}
\section{Inductive Relational GNN for Recommendation} 
\label{sec:metodology}
\begin{figure*}
    \centering
    \includegraphics[width=\textwidth, trim=0 565 140 100]{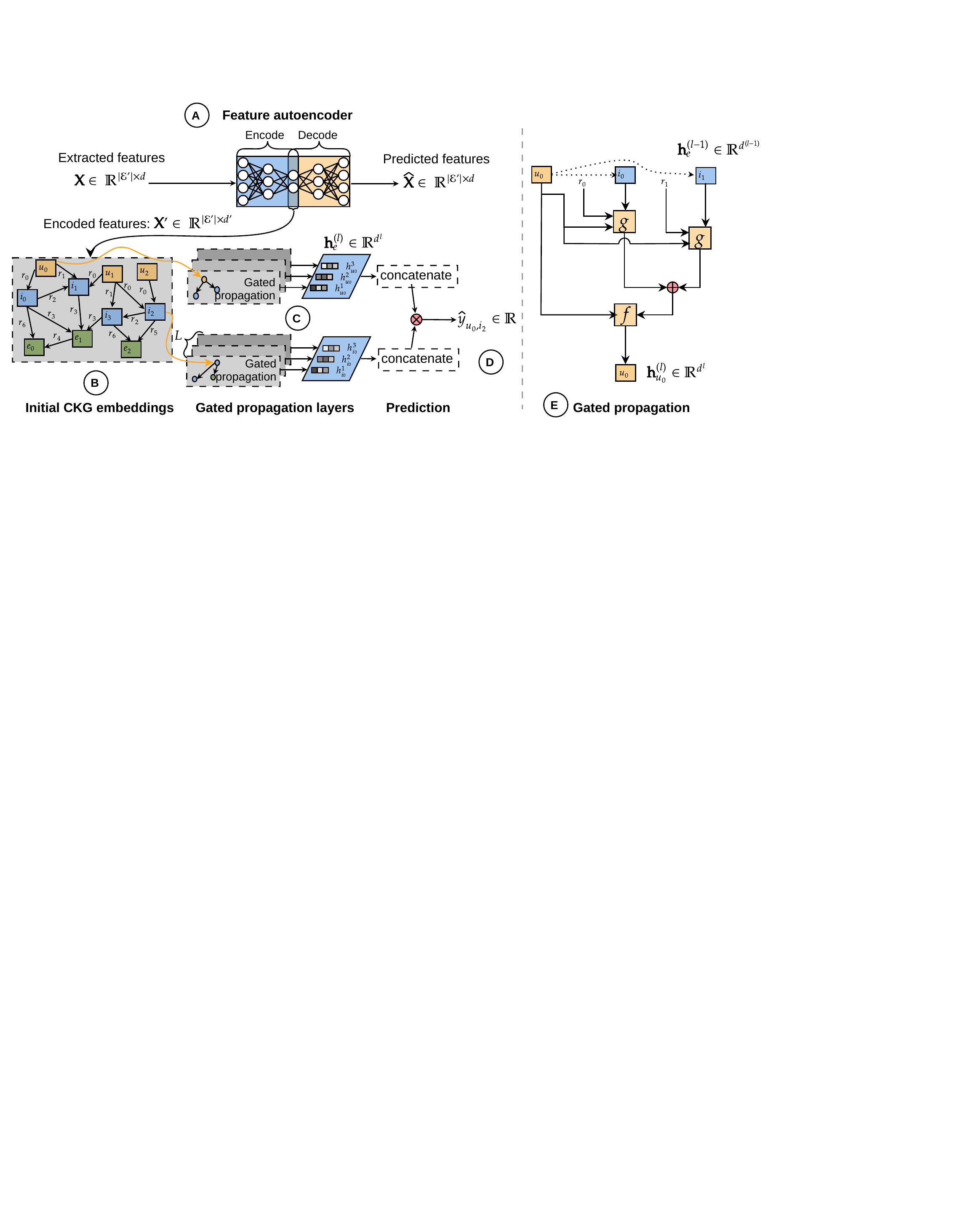}
    \caption{Illustration of the model.}
    \label{fig:model}
\end{figure*}

We now present our architecture \textit{\acrfull{own}}, which is able to generate node vector representations for recommendation in an inductive manner for both users and items using a \gls{ckg} without any user metadata.
The model consists of three components: 
\begin{enumerate*}[label=(\roman*)]
    \item an embedding layer, where we compress node feature information to create node embeddings (see \autoref{fig:model} A); 
    \item a gated propagation layer, which chooses which information to propagate from the embeddings of neighboring nodes in a \gls{ckg} (see \autoref{fig:model} C); 
    and
    \item a prediction layer, which uses the output of all propagation layers to create a user and an item embedding, producing a ranking score (see \autoref{fig:model} D).
\end{enumerate*}

\para{Embedding.} 
As for other methods exploiting KGs~\cite{hamilton2017inductive,ying2018graphpinsage}, we assume most items and connected entities have textual information and employ node embedding techniques and node degrees as initial features.
We use a pre-trained version of Sentence-BERT~\cite{reimers2019sentencebert} to process the textual description of each entity and produce sentence embeddings.
Nonetheless, when the non-recommendable entities $\entities$ are missing textual information, we can still 
learn their embeddings using \gls{kge} methods,  
since these entities are static (i.e., the KG excluding user and items, is not bound to change often).
Specifically, we use ComplEX~\cite{trouillon2016complex}, trained with all triples in the \gls{kg}, where we limit the output embeddings to be only on connected entities $\entities$, while items $\recs$ are always represented using textual descriptions.
\emph{Thus, we do not need retraining if new users or items are added.}
Yet, the initial features are too large ($d{\ge}756$) for the subsequent layers leading to out of memory errors and very slow training times.
The dimensionality therefore has to be reduced before being passed to the \gls{gnn} layers.
Hence, we introduce an \gls{ae} layer to reduce dimensionality~\cite{kramer1991nonlinearautoencoder}.
The loss of the \gls{ae} is defined as:

\vspace*{-8pt}
\begin{equation}\label{eqn:autoencoder_loss}
    \loss_{AE} = \mse\Bigl({\iembedding}, {\text{AE}}_{\text{de}}\bigl(\text{AE}_{\text{en}}({\iembedding})\bigr)\Bigr)
\end{equation}

{\noindent}where $\text{AE}_{\text{en}}$:$ \mathcal{R}^{|\entities'| {\times}d}{\mapsto}\mathcal{R}^{|\entities'| {\times}d'}$, with $d'{\ll}d$, is the \emph{encoding} function mapping the initial feature vector for each node to a set of lower dimensionality vectors.
We define an \gls{ae} for each embedding type, i.e., one for textual embeddings and one for \gls{kge}, converting the embeddings into a shared embedding space. The output of the encoder is used as the input for the \glspl{gnn} (see \autoref{fig:model} B).

\para{Propagation.} 
We apply multiple layers of \glspl{gnn}~\cite{hamilton2017inductive,ying2018graphpinsage}.
Yet, existing inductive GNN are not designed to use relational information.
We therefore define relational gates 
as:
\begin{equation}
    \hidden{\mathcal{N}_\head}^{(l+1)} = \frac{1}{|\mathcal{F}_N(\head)|} \sum_{(\head, \relation, \tail) \in \mathcal{F}_N(\head)} g^{(l+1)}(\head, \relation, \tail) \hidden{\tail}^{(l)}
\end{equation}
where $\mathcal{F}_N(\head)$ defines the ego-network of node $\head{\in}\verts$, $g$ is the gating function, and $\hidden{\tail}^{(l)}{\in}\mathbb{R}^{d^{(l)}}$ is the embedding of node $\tail{\in}\verts$. 
During the first graph convolution $d^{(0)}{=}d'$, i.e., the output dimensionality of the \gls{ae}.
In contrast to other gated networks (e.g., MGAT~\cite{tao2020mgat} and GGCN~\cite{li2015gatedgcnggnn}), our model's gates are relation-specific, allowing it to propagate different information from different parts of an entity's embedding based on the relation to it\footnote{See \autoref{app:abl} for the gating mechanism related improvements.}.
Hence, our gate is $g^{(l+1)}(\head, \relation, \tail){=} \sigma\bigl(\cmatrix{W}^{(l+1)}_{\relation}(\hidden{\head}^{(l)}\|\hidden{\tail}^{(l)})\bigr)$, with the sigmoid activation function $\sigma$, the concatenation operator $\|$, and $\cmatrix{W}^{(l+1)}_{\relation}\in\mathbb{R}^{d^{(l)}\times 2d^{(l)}}$.

The final part combines an entity's current embedding $\hidden{\head}^{(l)}$ with the aggregated ego-network embedding $\hidden{\mathcal{N}_\head}^{(l+1)}$, defined as $\hidden{\head}^{(l+1)}{=}f(\hidden{\head}^{(l)}, \hidden{\mathcal{N}_\head}^{(l+1)})$, where $f$ is an aggregator function. 
Among recent aggregators~\cite{kipf2016semigcn,wang2019kgat,hamilton2017inductive}, we identified the best performance with 
$f(\hidden{\head}^{(l)}, \hidden{\mathcal{N}_\head}^{(l+1)}) = \hidden{\mathcal{N}_\head}^{(l+1)}$~\cite{he2020lightgcn}. We refer to \autoref{app:agg} for the formal definition of the aggregators. 

\para{Prediction.} As in previous approaches~\cite{kipf2016semigcn,wang2019kgat,tao2020mgat}, we concatenate the output after each layer for a user $\user$ and item $\rec$ as:
$
    \hidden{\user}^* = \hidden{\user}^1 \| \ldots \| \hidden{\user}^L,$ and  $\hidden{\rec}^* = \hidden{\rec}^1 \| \ldots \| \hidden{\rec}^L
$
efficiently obtaining the final prediction via dot-product as $\hat{y}_{\user\rec}{=}\hidden{\user}^{*\top}{\hidden{\rec}^*}$, which usually outperforms learned, non-linear, similarities~\cite{rendle2020neuralvsfactorization}.
    

\para{Optimization.} The final loss function is a combination of autoencoder loss in \autoref{eqn:autoencoder_loss} and the \gls{bpr} loss~\cite{wang2019kgat,he2020lightgcn,tao2020mgat,rendle2012bpr}, s.t. we learn to encode an embedding suitable for ranking while maintaining the information of the original features, computed as:
    ${\loss} = {\loss}_{BPR} + \lambda {\loss}_{AE} + \gamma \|\Theta\|_2^2$,
where $\lambda$ and $\gamma$ are tuned during hyperparameter optimization and $\Theta$ are the learnable parameters. We study the effects of the \gls{ae} loss in \autoref{app:abl}.

\begin{table}[bt]
\centering
\caption{Dataset statistics}
\vspace*{-12pt}
\resizebox{1\linewidth}{!}{%
\begin{tabular}{l|r|r|RRRr|r|}
 & \multicolumn{1}{c|}{\textbf{\acrshort{ml}}} & \multicolumn{1}{c|}{\textbf{\acrshort{mls}}} & \multicolumn{1}{R}{\textbf{synthetic}} & \multicolumn{1}{R}{\textbf{\acrshort{syn}}} & \multicolumn{1}{R}{\textbf{\acrshort{mr}}} & \multicolumn{1}{c|}{\textbf{\acrshort{ab}}} & \multicolumn{1}{c|}{\textbf{\acrshort{abs}}} \\ \hline
\textbf{\# Users}   & 132,287     & 12,500      & 100,000     & 12,500     & 2820   & 70,679  & 60,000\\ 
\textbf{\# Items}   & 4725        & 4438        & 4735        & 4724       & 1307   & 24,841  & 24,841\\ 
\textbf{\# Ratings} & 11,376,533  & 1,106,000   & 194,670,328 & 24,237,937 & 38,971 & 847,733 & 720,111\\ 
\textbf{Density}    & 0.018      & 0.019      & 0.0355      & 0.1634     & 0.0046 & 0.00048 & 0.00048\\ \hline
\end{tabular}
}
\label{table:datasets}
\vspace*{-12pt}
\end{table}
\begin{table}[bt]
\caption{Knowledge graph statistics}
\label{table:kgstats}
\vspace*{-12pt}
\resizebox{1\linewidth}{!}{%
\begin{tabular}{l|r|r|r|r|}
            & \multicolumn{1}{c|}{\textbf{Entities}} & \multicolumn{1}{c|}{\textbf{Edges}} & \multicolumn{1}{c|}{\textbf{Labels}} & \multicolumn{1}{c|}{\textbf{Density}} \\ \hline
\textbf{MindReader~\cite{brams2020mindreader}} & 13,767 & 201,438   & 8  & 1.06e-3 \\
\textbf{Amazon Book~\cite{wang2019kgat}}          & 88,572 & 2,555,995 & 39 & 3.26e-4 \\ \hline
\end{tabular}
}
\vspace*{-12pt}
\end{table}

\para{Scalability.} 
Our embedding approach, calculating the aggregation ($\hidden{h}^{l+1}$), and the prediction ($\hat{y}$) are all bounded by the number of nodes in the graph, while the calculation of the ego-network ($ \hidden{\mathcal{N}_\head}^{(l+1)}$) is bounded by the number of edges. 
As these steps are applied sequentially and $|\verts|{\ll}|\edges|$, we know that the complexity of our method is bounded by the ego-network aggregation complexity, more specifically, the linear transformation of the gate calculation.
When na\"ively applying the gates over all edges, the complexity is $O(|\edges|d)$, where $d$ is the largest dimension utilized during graph convolutions\footnote{We note that $|\edges|$ is bounded by $O(|\verts|^2|\relations|)$.}.
Yet, as $\cmatrix{W}_{\relation}(\hidden{\head}\|\hidden{\tail})$ is equivalent to $\cmatrix{W}^1_{\relation}\hidden{\head}+\cmatrix{W}^2_{\relation}\hidden{\tail}$ we only need to compute the transformation for each unique $(\head, \relation)$ and $(\relation, \tail)$ pair instead of each unique $(\head, \relation, \tail)$ triple. 
Therefore, we can apply a MapReduce computation~\cite{ying2018graphpinsage} to have at most $2|\verts||\relations|$ calculations, leading to the complexity $O(|\verts||\relations|d){\ll}O(|\edges|d)$.

As our prediction is a dot product after graph convolutions, our method can predict in $O(|\hidden{\user}^*|\cdot |\recs|)$ for a single user as the vector dot product complexity is $O(|\hidden{\user}^*|)$, which we do $|\recs|$ number of times. 
This is less than PinSAGE, which requires the dot product between all items and the user's rated items, s.t. $\left(\frac{|\recs|}{2}\right)^2=O(|\recs|^2)$ for a worst-case where a user has rated half the entities and we want to rank the other half. Our prediction is less computationally complex, as the reduced dimension, even when taking the number of layers into account as $d*L\ll |\recs|$, is far less than the number of items.


\vspace*{-7pt}
\section{Experiments}
\label{sec:experiments}
Inductive approaches are designed to provide recommendations in a cold-start setting, where ratings for new users are only known at inference time. 
Yet, as we will show, these baselines do not perform in this setting due to poor selection of learning metrics, evaluation methodologies, or other complexities. 
In the following, we aim at answering the questions:
\begin{enumerate*}[label=\bfseries RQ\arabic*)]
    \item Are existing inductive approaches really competitive when compared to \gls{own}? 
    \item Why do other architectures under-perform? and,
    \item What negative item sampling strategy is the most appropriate for the evaluation?
\end{enumerate*}

\begin{table*}[!htb]
\centering
\caption{Results on the different dataset. `*' represents statistical significant increase over the best baseline. NDCG, Recall and Precision is performed at 20, while I-NDCG is the calculation method used in \cite{wu2021towardsidcf} for the full subsampled list.}
\label{table:results}
\vspace*{-12pt}
\resizebox{1\textwidth}{!}{%
\begin{tabular}{l|rrrRr|rrrRr|rrrRr}
 & \multicolumn{5}{c|}{\textbf{MovieLens Subsampled + 1250 users}} & \multicolumn{5}{c|}{\textbf{MovieLens Subsampled + 90\%}} & \multicolumn{5}{c}{\textbf{Amazon Book Subsampled + 15\%}} \\ \hline
 & \multicolumn{1}{c}{\textbf{NDCG}} & \multicolumn{1}{c}{\textbf{Recall}} & \multicolumn{1}{c}{\textbf{Precision}} & \multicolumn{1}{R}{\textbf{Cov}} & \multicolumn{1}{c|}{\textbf{I-NDCG}} & \multicolumn{1}{c}{\textbf{NDCG}} & \multicolumn{1}{c}{\textbf{Recall}} & \multicolumn{1}{c}{\textbf{Precision}} & \multicolumn{1}{R}{\textbf{Cov}} & \multicolumn{1}{c|}{\textbf{I-NDCG}} & \multicolumn{1}{c}{\textbf{NDCG}} & \multicolumn{1}{c}{\textbf{Recall}} & \multicolumn{1}{c}{\textbf{Precision}} & \multicolumn{1}{R}{\textbf{Cov}} & \multicolumn{1}{c}{\textbf{I-NDCG}} \\ \hline
\textbf{TopPop} 
  & 0.11182 & 0.14351 & 0.05256 & 0.01587 & 0.85136 
  & 0.10916 & 0.14426 & 0.05147 & 0.02222 & 0.85350 
  & 0.01552 & 0.03496 & 0.00273 & 0.00125 & 0.78715 \\ \hline
\textbf{GraphSAGE} 
  & 0.07862 & 0.10637 & 0.04336 & \textbf{0.06307} & 0.84262 
  & 0.07281 & 0.10356 & 0.04164 & \textbf{0.31471} & 0.84345 
  & 0.00667 & 0.01695 & 0.00149 & \textbf{0.14243} & 0.76845 \\ \hline
\textbf{PinSAGE} 
  & \textbf{0.14129} & \textbf{0.18745} & \textbf{0.06396} & 0.02857 & \textbf{0.88378} 
  & \textbf{0.13592} & \textbf{0.18622} & \textbf{0.06280} & 0.04169 & \textbf{0.88496} 
  & \textbf{0.05398} & \textbf{0.11764} & \textbf{0.00927} & 0.12439 & \textbf{0.89937} \\ \hline
\textbf{IDCF} 
  & 0.11042 & 0.14155 & 0.05244 & 0.01566 & 0.81959 
  & 0.10875 & 0.14394 & 0.05113 & 0.02201 & 0.85388 
  & 0.01205 & 0.02828 & 0.00231 & 0.00845 & 0.75548 \\ \hline\hline
\textbf{SimpleRec} 
  & \textbf{0.18085*} & \textbf{0.24618*} & \textbf{0.08460*} & \textbf{0.18540*} & \textbf{0.92200*} 
  & \textbf{0.18411*} & \textbf{0.24482*} & \textbf{0.08365*} & \textbf{0.42646*} & \textbf{0.92208*} 
  & \textbf{0.06472*} & \textbf{0.14055*} & \textbf{0.01144*} & \textbf{0.21460*} & \textbf{0.91769*} \\ \hline
\end{tabular}%
}
\vspace*{-12pt}
\end{table*}

\para{Datasets.}
To evaluate existing methods, we adopt two real world datasets: 
\begin{enumerate*}[label=(\roman*)]
\item \gls{ml}~\cite{harper2015movielens}, a dataset with ratings on movies and
\item \gls{ab}~\cite{ni2019justifyingamazonreviewdataset}, a dataset with reviews on books.
\end{enumerate*}
Neither dataset has an associated \gls{kg}. 
We therefore use the MindReader \gls{kg}~\cite{brams2020mindreader} for the \gls{ml} dataset, and  
for the \gls{ab} dataset the KG constructed to evaluate KGAT~\cite{wang2019kgat}. 
In both cases, we keep only items mapped to the \gls{kg}. 
We highlight how some methods compare performances over different versions of the \gls{ab} dataset, e.g., the 2018 used in IDCF~\cite{wu2021towardsidcf} is different from the version used in other works~\cite{wang2019kgat,he2020lightgcn,wang2019neuralngcf,bars2022datasets}.
Since users and items have an average of $39.9$ and $22.9$ ratings, respectively, on the 2018 dataset, while the users have an average of $12.0$ ratings and items $34.1$ in the 2014 dataset used here, \emph{our evaluation considers a sparse dataset, which is a more challenging setup}.
Therefore, comparisons on just the original reported numbers are unreliable.

In our cold-start experiments, we sample 12,500 users from \gls{ml} and 60,000 users from \gls{ab} for training, named \gls{mls} and \gls{abs}, respectively. 
We then created two cold-start scenarios on \gls{mls}: one adding 10\% new users (i.e., 1250); and an extreme setting,  where we treat all users not in \gls{mls} as cold-start users, being ${\sim}90\%$ of the users in the original \gls{ml} dataset. 
For \gls{abs} we create one scenario adding the remaining users from the original dataset, being ${\sim}15\%$ of total number of users.

\para{Parameter settings.} 
All models are implemented in PyTorch and optimized using the Adam optimizer\footnote{We tested the methods' implementations also on the datasets reported in their evaluation, when possible, getting similar results.}; we train with a batch size of 1024 for a maximal of 1000 epochs, saving the best-performing state on the validation set and stopping after 50 successive epochs without improvement.
For hyperparameter tuning, we employ \gls{asha}~\cite{li2018massivelyasha}, with $r=4$, $R=256$, $\eta=4$, setting minimum and maximum values of parameters based on the 
values described original works\footnote{A configuration file can be found in the source code with all parameter options.}.

\para{Methods.} 
We compare to four methods: TopPop~\cite{cremonesi2010topnperformance}, a simple non-personalized recommender that recommends the most popular items (a common baseline~\cite{palumbo2020entity2rec}); GraphSAGE~\cite{hamilton2017inductive}, modified to recommend using the cosine similarity between a user's rated items and new items; PinSAGE\cite{ying2018graphpinsage}, with a semi-supervised objective, i.e., items co-rated should be similar, closely matching their pin/boards setting\footnote{Based on \url{https://github.com/dmlc/dgl/tree/master/examples/pytorch/pinsage}}; and IDCF~\cite{wu2021towardsidcf}, a two-step learning method, which reports experimental results superior to all other inductive methods to date.

\para{RQ1 \& RQ2.} 
As \autoref{table:results} shows, \gls{own} is able to outperform all methods on all metrics with statistical significance. 
We also see contrasting results w.r.t. the original IDCF evaluation.
IDCF was originally evaluated in a ranking setting; however, the learned embeddings of IDCF are learned towards Cross-Entropy, a non-ranking, point-wise learning objective.
Such learning methodologies have been shown to perform poorly, learning a similar or worse ranking than TopPop~\cite{cremonesi2010topnperformance,wu2022effectivenessbprbce}.
In the original work, IDCF outperforms PinSAGE by a small margin, yet we observe the opposite in our evaluation.
This is due to (i) our setup performing a more appropriate early stopping based on the evaluation metric instead of the loss function and (ii) our evaluation adopting a better learning objective for PinSAGE.
Ranking loss generally leads to better performance, utilized by the two best performing methods.
Yet, in IDCF method use the Cross-Entropy loss~\cite{walid2020evaluationmetrics}.

\para{RQ3.}
When evaluating ranking performance, NDCG is the metric commonly adopted, but there are two alternatives on which set of items to rank: either rank all items in the the dataset or just a subset.
In other evaluations~\cite{wu2021towardsidcf,he2017neuralncf,cremonesi2010topnperformance}, instead of ranking all items, only $N$ negative items are sampled per each positive. 
While this
makes it equally hard to rank positive items for all test users,
it has been proven to produce unreliable comparisons of performances across methods~\cite{walid2020evaluationmetrics}.
\emph{In the original IDCF evaluation, this faulty method (here labeled I-NDCG) is adopted.}
NDCG measures the methods' ability to provide high-quality recommendations in the top of a ranked list.
Yet, when subsampling negative items, the methods' relative performance can change, meaning that the best performing method might not provide \textit{any} or few relevant recommendations when challenged with the complete list of possible items.
E.g., we see GraphSAGE outperforms IDCF on I-NDCG in both \gls{mls}${+}1250$ and \gls{abs}${+}15\%$, though clearly performing worse in the appropriate NDCG@20.
On a different Amazon dataset, the Amazon-Beauty dataset, not reported here for space constraints, when utilizing the faulty I-NDCG, we also found TopPop (not included in the IDCF evaluation) outperforming \textit{all} methods results (including IDCF) reported in IDCF~\cite{wu2021towardsidcf}.
However, when using NDCG@20 we see up towards 3x times better performance when utilizing our method, compared to TopPop. 

\vspace*{-7pt}

\section{Conclusion and future work}
\label{sec:conclusion}
In this work, we propose a simple and strong baseline for inductive recommendation.
Our goal was to demonstrate the feasibility and advantages of this kind of architecture when compared to other inductive architectures.
In particular, we devise a scalable GNN architecture to perform inductive learning for recommendation, able to utilize high-order connectivities in \glspl{ckg}.
We show that our method can outperform related work and showcase methodological limitations in the evaluation methods used.
Overall, we conclude that more attention is needed towards this kind of architecture, especially given their ability to: (1) scale to large graphs and large numbers of users using MapReduce, and (2) maintain good prediction while being simple; i.e, performing no complex sampling methods or non-standard learning steps. 
Finally, we raise attention to the need for more sound evaluation protocols and for more transparency on the dataset adopted.



\section*{Acknowledgements}
This research was partially funded by the Danish Council for Independent Research (DFF) under grant agreement no. DFF-8048-00051B and the Poul Due Jensen Foundation. During this work, Matteo Lissandrini was supported by the European Union’s Horizon 2020 research and innovation programme under the Marie Skłodowska-Curie grant agreement No 838216.

\bibliographystyle{ACM-Reference-Format}
\bibliography{references}


\begin{thebibliography}{35}


\ifx \showCODEN    \undefined \def \showCODEN     #1{\unskip}     \fi
\ifx \showDOI      \undefined \def \showDOI       #1{#1}\fi
\ifx \showISBNx    \undefined \def \showISBNx     #1{\unskip}     \fi
\ifx \showISBNxiii \undefined \def \showISBNxiii  #1{\unskip}     \fi
\ifx \showISSN     \undefined \def \showISSN      #1{\unskip}     \fi
\ifx \showLCCN     \undefined \def \showLCCN      #1{\unskip}     \fi
\ifx \shownote     \undefined \def \shownote      #1{#1}          \fi
\ifx \showarticletitle \undefined \def \showarticletitle #1{#1}   \fi
\ifx \showURL      \undefined \def \showURL       {\relax}        \fi
\providecommand\bibfield[2]{#2}
\providecommand\bibinfo[2]{#2}
\providecommand\natexlab[1]{#1}
\providecommand\showeprint[2][]{arXiv:#2}

\bibitem[\protect\citeauthoryear{Brams, Jakobsen, Jendal, Lissandrini, Dolog,
  and Hose}{Brams et~al\mbox{.}}{2020}]%
        {brams2020mindreader}
\bibfield{author}{\bibinfo{person}{Anders~H. Brams}, \bibinfo{person}{Anders~L.
  Jakobsen}, \bibinfo{person}{Theis~E. Jendal}, \bibinfo{person}{Matteo
  Lissandrini}, \bibinfo{person}{Peter Dolog}, {and} \bibinfo{person}{Katja
  Hose}.} \bibinfo{year}{2020}\natexlab{}.
\newblock \showarticletitle{MindReader: Recommendation over Knowledge Graph
  Entities with Explicit User Ratings}. In
  \bibinfo{booktitle}{\emph{Proceedings of the 29th ACM International
  Conference on Information \& Knowledge Management}} (Virtual Event, Ireland)
  \emph{(\bibinfo{series}{CIKM '20})}. \bibinfo{publisher}{Association for
  Computing Machinery}, \bibinfo{address}{New York, NY, USA},
  \bibinfo{pages}{2975–2982}.
\newblock
\showISBNx{9781450368599}
\urldef\tempurl%
\url{https://doi.org/10.1145/3340531.3412759}
\showDOI{\tempurl}


\bibitem[\protect\citeauthoryear{Cremonesi, Koren, and Turrin}{Cremonesi
  et~al\mbox{.}}{2010}]%
        {cremonesi2010topnperformance}
\bibfield{author}{\bibinfo{person}{Paolo Cremonesi}, \bibinfo{person}{Yehuda
  Koren}, {and} \bibinfo{person}{Roberto Turrin}.}
  \bibinfo{year}{2010}\natexlab{}.
\newblock \showarticletitle{Performance of recommender algorithms on top-n
  recommendation tasks}. In \bibinfo{booktitle}{\emph{Proceedings of the fourth
  ACM conference on Recommender systems}}. \bibinfo{pages}{39--46}.
\newblock


\bibitem[\protect\citeauthoryear{for Recommender~Systems}{for
  Recommender~Systems}{2022}]%
        {bars2022datasets}
\bibfield{author}{\bibinfo{person}{BenchmArking for Recommender~Systems}.}
  \bibinfo{year}{2022}\natexlab{}.
\newblock \bibinfo{booktitle}{\emph{Bars Datasets}}.
\newblock
\urldef\tempurl%
\url{https://github.com/openbenchmark/BARS/tree/master/candidate_matching/datasets}
\showURL{%
\tempurl}


\bibitem[\protect\citeauthoryear{Hamilton, Ying, and Leskovec}{Hamilton
  et~al\mbox{.}}{2017}]%
        {hamilton2017inductive}
\bibfield{author}{\bibinfo{person}{William~L Hamilton}, \bibinfo{person}{Rex
  Ying}, {and} \bibinfo{person}{Jure Leskovec}.}
  \bibinfo{year}{2017}\natexlab{}.
\newblock \showarticletitle{Inductive representation learning on large graphs}.
  In \bibinfo{booktitle}{\emph{Proceedings of the 31st International Conference
  on Neural Information Processing Systems}}. \bibinfo{pages}{1025--1035}.
\newblock


\bibitem[\protect\citeauthoryear{Harper and Konstan}{Harper and
  Konstan}{2015}]%
        {harper2015movielens}
\bibfield{author}{\bibinfo{person}{F~Maxwell Harper} {and}
  \bibinfo{person}{Joseph~A Konstan}.} \bibinfo{year}{2015}\natexlab{}.
\newblock \showarticletitle{The movielens datasets: History and context}.
\newblock \bibinfo{journal}{\emph{Acm transactions on interactive intelligent
  systems (tiis)}} \bibinfo{volume}{5}, \bibinfo{number}{4}
  (\bibinfo{year}{2015}), \bibinfo{pages}{1--19}.
\newblock


\bibitem[\protect\citeauthoryear{He, Deng, Wang, Li, Zhang, and Wang}{He
  et~al\mbox{.}}{2020}]%
        {he2020lightgcn}
\bibfield{author}{\bibinfo{person}{Xiangnan He}, \bibinfo{person}{Kuan Deng},
  \bibinfo{person}{Xiang Wang}, \bibinfo{person}{Yan Li},
  \bibinfo{person}{Yongdong Zhang}, {and} \bibinfo{person}{Meng Wang}.}
  \bibinfo{year}{2020}\natexlab{}.
\newblock \showarticletitle{Lightgcn: Simplifying and powering graph
  convolution network for recommendation}. In
  \bibinfo{booktitle}{\emph{Proceedings of the 43rd International ACM SIGIR
  Conference on Research and Development in Information Retrieval}}.
  \bibinfo{pages}{639--648}.
\newblock


\bibitem[\protect\citeauthoryear{He, Liao, Zhang, Nie, Hu, and Chua}{He
  et~al\mbox{.}}{2017}]%
        {he2017neuralncf}
\bibfield{author}{\bibinfo{person}{Xiangnan He}, \bibinfo{person}{Lizi Liao},
  \bibinfo{person}{Hanwang Zhang}, \bibinfo{person}{Liqiang Nie},
  \bibinfo{person}{Xia Hu}, {and} \bibinfo{person}{Tat-Seng Chua}.}
  \bibinfo{year}{2017}\natexlab{}.
\newblock \showarticletitle{Neural collaborative filtering}. In
  \bibinfo{booktitle}{\emph{Proceedings of the 26th international conference on
  world wide web}}. \bibinfo{pages}{173--182}.
\newblock


\bibitem[\protect\citeauthoryear{Jain and Dhillon}{Jain and Dhillon}{2013}]%
        {jain2013provableimc}
\bibfield{author}{\bibinfo{person}{Prateek Jain} {and}
  \bibinfo{person}{Inderjit~S Dhillon}.} \bibinfo{year}{2013}\natexlab{}.
\newblock \showarticletitle{Provable inductive matrix completion}.
\newblock \bibinfo{journal}{\emph{arXiv preprint arXiv:1306.0626}}
  (\bibinfo{year}{2013}).
\newblock


\bibitem[\protect\citeauthoryear{Kipf and Welling}{Kipf and Welling}{2016}]%
        {kipf2016semigcn}
\bibfield{author}{\bibinfo{person}{Thomas~N Kipf} {and} \bibinfo{person}{Max
  Welling}.} \bibinfo{year}{2016}\natexlab{}.
\newblock \showarticletitle{Semi-supervised classification with graph
  convolutional networks}.
\newblock \bibinfo{journal}{\emph{arXiv preprint arXiv:1609.02907}}
  (\bibinfo{year}{2016}).
\newblock


\bibitem[\protect\citeauthoryear{Kramer}{Kramer}{1991}]%
        {kramer1991nonlinearautoencoder}
\bibfield{author}{\bibinfo{person}{Mark~A Kramer}.}
  \bibinfo{year}{1991}\natexlab{}.
\newblock \showarticletitle{Nonlinear principal component analysis using
  autoassociative neural networks}.
\newblock \bibinfo{journal}{\emph{AIChE journal}} \bibinfo{volume}{37},
  \bibinfo{number}{2} (\bibinfo{year}{1991}), \bibinfo{pages}{233--243}.
\newblock


\bibitem[\protect\citeauthoryear{Krichene and Rendle}{Krichene and
  Rendle}{2020}]%
        {walid2020evaluationmetrics}
\bibfield{author}{\bibinfo{person}{Walid Krichene} {and}
  \bibinfo{person}{Steffen Rendle}.} \bibinfo{year}{2020}\natexlab{}.
\newblock \showarticletitle{On Sampled Metrics for Item Recommendation}. In
  \bibinfo{booktitle}{\emph{Proceedings of the 26th ACM SIGKDD International
  Conference on Knowledge Discovery and Data Mining}} (Virtual Event, CA, USA)
  \emph{(\bibinfo{series}{KDD '20})}. \bibinfo{publisher}{Association for
  Computing Machinery}, \bibinfo{address}{New York, NY, USA},
  \bibinfo{pages}{1748–1757}.
\newblock
\showISBNx{9781450379984}
\urldef\tempurl%
\url{https://doi.org/10.1145/3394486.3403226}
\showDOI{\tempurl}


\bibitem[\protect\citeauthoryear{Lee, Im, Jang, Cho, and Chung}{Lee
  et~al\mbox{.}}{2019}]%
        {lee2019melu}
\bibfield{author}{\bibinfo{person}{Hoyeop Lee}, \bibinfo{person}{Jinbae Im},
  \bibinfo{person}{Seongwon Jang}, \bibinfo{person}{Hyunsouk Cho}, {and}
  \bibinfo{person}{Sehee Chung}.} \bibinfo{year}{2019}\natexlab{}.
\newblock \showarticletitle{Melu: Meta-learned user preference estimator for
  cold-start recommendation}. In \bibinfo{booktitle}{\emph{Proceedings of the
  25th ACM SIGKDD International Conference on Knowledge Discovery \& Data
  Mining}}. \bibinfo{pages}{1073--1082}.
\newblock


\bibitem[\protect\citeauthoryear{Li, Jamieson, Rostamizadeh, Gonina, Hardt,
  Recht, and Talwalkar}{Li et~al\mbox{.}}{2018}]%
        {li2018massivelyasha}
\bibfield{author}{\bibinfo{person}{Liam Li}, \bibinfo{person}{Kevin Jamieson},
  \bibinfo{person}{Afshin Rostamizadeh}, \bibinfo{person}{Ekaterina Gonina},
  \bibinfo{person}{Moritz Hardt}, \bibinfo{person}{Ben Recht}, {and}
  \bibinfo{person}{Ameet Talwalkar}.} \bibinfo{year}{2018}\natexlab{}.
\newblock \showarticletitle{Massively parallel hyperparameter tuning}.
\newblock  (\bibinfo{year}{2018}).
\newblock


\bibitem[\protect\citeauthoryear{Li, Tarlow, Brockschmidt, and Zemel}{Li
  et~al\mbox{.}}{2016}]%
        {li2015gatedgcnggnn}
\bibfield{author}{\bibinfo{person}{Yujia Li}, \bibinfo{person}{Daniel Tarlow},
  \bibinfo{person}{Marc Brockschmidt}, {and} \bibinfo{person}{Richard~S.
  Zemel}.} \bibinfo{year}{2016}\natexlab{}.
\newblock \showarticletitle{Gated Graph Sequence Neural Networks}. In
  \bibinfo{booktitle}{\emph{4th International Conference on Learning
  Representations, {ICLR} 2016, San Juan, Puerto Rico, May 2-4, 2016,
  Conference Track Proceedings}}, \bibfield{editor}{\bibinfo{person}{Yoshua
  Bengio} {and} \bibinfo{person}{Yann LeCun}} (Eds.).
\newblock
\urldef\tempurl%
\url{http://arxiv.org/abs/1511.05493}
\showURL{%
\tempurl}


\bibitem[\protect\citeauthoryear{Ni, Li, and McAuley}{Ni et~al\mbox{.}}{2019}]%
        {ni2019justifyingamazonreviewdataset}
\bibfield{author}{\bibinfo{person}{Jianmo Ni}, \bibinfo{person}{Jiacheng Li},
  {and} \bibinfo{person}{Julian McAuley}.} \bibinfo{year}{2019}\natexlab{}.
\newblock \showarticletitle{Justifying Recommendations using Distantly-Labeled
  Reviews and Fine-Grained Aspects}. In \bibinfo{booktitle}{\emph{Proceedings
  of the 2019 Conference on Empirical Methods in Natural Language Processing
  and the 9th International Joint Conference on Natural Language Processing
  (EMNLP-IJCNLP)}}. \bibinfo{publisher}{Association for Computational
  Linguistics}, \bibinfo{address}{Hong Kong, China}, \bibinfo{pages}{188--197}.
\newblock
\urldef\tempurl%
\url{https://doi.org/10.18653/v1/D19-1018}
\showDOI{\tempurl}


\bibitem[\protect\citeauthoryear{Palumbo, Monti, Rizzo, Troncy, and
  Baralis}{Palumbo et~al\mbox{.}}{2020}]%
        {palumbo2020entity2rec}
\bibfield{author}{\bibinfo{person}{Enrico Palumbo}, \bibinfo{person}{Diego
  Monti}, \bibinfo{person}{Giuseppe Rizzo}, \bibinfo{person}{Rapha{\"e}l
  Troncy}, {and} \bibinfo{person}{Elena Baralis}.}
  \bibinfo{year}{2020}\natexlab{}.
\newblock \showarticletitle{entity2rec: Property-specific knowledge graph
  embeddings for item recommendation}.
\newblock \bibinfo{journal}{\emph{Expert Systems with Applications}}
  \bibinfo{volume}{151} (\bibinfo{year}{2020}), \bibinfo{pages}{113235}.
\newblock


\bibitem[\protect\citeauthoryear{Reimers and Gurevych}{Reimers and
  Gurevych}{2019}]%
        {reimers2019sentencebert}
\bibfield{author}{\bibinfo{person}{Nils Reimers} {and} \bibinfo{person}{Iryna
  Gurevych}.} \bibinfo{year}{2019}\natexlab{}.
\newblock \showarticletitle{Sentence-BERT: Sentence Embeddings using Siamese
  BERT-Networks}. In \bibinfo{booktitle}{\emph{Proceedings of the 2019
  Conference on Empirical Methods in Natural Language Processing and the 9th
  International Joint Conference on Natural Language Processing
  (EMNLP-IJCNLP)}}. \bibinfo{pages}{3982--3992}.
\newblock


\bibitem[\protect\citeauthoryear{Rendle, Freudenthaler, Gantner, and
  Schmidt-Thieme}{Rendle et~al\mbox{.}}{2009}]%
        {rendle2012bpr}
\bibfield{author}{\bibinfo{person}{Steffen Rendle}, \bibinfo{person}{Christoph
  Freudenthaler}, \bibinfo{person}{Zeno Gantner}, {and} \bibinfo{person}{Lars
  Schmidt-Thieme}.} \bibinfo{year}{2009}\natexlab{}.
\newblock \showarticletitle{BPR: Bayesian personalized ranking from implicit
  feedback}. In \bibinfo{booktitle}{\emph{Proceedings of the Twenty-Fifth
  Conference on Uncertainty in Artificial Intelligence}}.
  \bibinfo{pages}{452--461}.
\newblock


\bibitem[\protect\citeauthoryear{Rendle, Krichene, Zhang, and Anderson}{Rendle
  et~al\mbox{.}}{2020}]%
        {rendle2020neuralvsfactorization}
\bibfield{author}{\bibinfo{person}{Steffen Rendle}, \bibinfo{person}{Walid
  Krichene}, \bibinfo{person}{Li Zhang}, {and} \bibinfo{person}{John
  Anderson}.} \bibinfo{year}{2020}\natexlab{}.
\newblock \showarticletitle{Neural collaborative filtering vs. matrix
  factorization revisited}. In \bibinfo{booktitle}{\emph{Fourteenth ACM
  Conference on Recommender Systems}}. \bibinfo{pages}{240--248}.
\newblock


\bibitem[\protect\citeauthoryear{Tao, Wei, Wang, He, Huang, and Chua}{Tao
  et~al\mbox{.}}{2020}]%
        {tao2020mgat}
\bibfield{author}{\bibinfo{person}{Zhulin Tao}, \bibinfo{person}{Yinwei Wei},
  \bibinfo{person}{Xiang Wang}, \bibinfo{person}{Xiangnan He},
  \bibinfo{person}{Xianglin Huang}, {and} \bibinfo{person}{Tat-Seng Chua}.}
  \bibinfo{year}{2020}\natexlab{}.
\newblock \showarticletitle{MGAT: Multimodal Graph Attention Network for
  Recommendation}.
\newblock \bibinfo{journal}{\emph{Information Processing \& Management}}
  \bibinfo{volume}{57}, \bibinfo{number}{5} (\bibinfo{year}{2020}),
  \bibinfo{pages}{102277}.
\newblock


\bibitem[\protect\citeauthoryear{Trouillon, Welbl, Riedel, Gaussier, and
  Bouchard}{Trouillon et~al\mbox{.}}{2016}]%
        {trouillon2016complex}
\bibfield{author}{\bibinfo{person}{Th{\'{e}}o Trouillon},
  \bibinfo{person}{Johannes Welbl}, \bibinfo{person}{Sebastian Riedel},
  \bibinfo{person}{{\'{E}}ric Gaussier}, {and} \bibinfo{person}{Guillaume
  Bouchard}.} \bibinfo{year}{2016}\natexlab{}.
\newblock \showarticletitle{Complex Embeddings for Simple Link Prediction}. In
  \bibinfo{booktitle}{\emph{Proceedings of the 33nd International Conference on
  Machine Learning, {ICML} 2016, New York City, NY, USA, June 19-24, 2016}}
  \emph{(\bibinfo{series}{{JMLR} Workshop and Conference Proceedings},
  Vol.~\bibinfo{volume}{48})},
  \bibfield{editor}{\bibinfo{person}{Maria{-}Florina Balcan} {and}
  \bibinfo{person}{Kilian~Q. Weinberger}} (Eds.).
  \bibinfo{publisher}{JMLR.org}, \bibinfo{pages}{2071--2080}.
\newblock
\urldef\tempurl%
\url{http://proceedings.mlr.press/v48/trouillon16.html}
\showURL{%
\tempurl}


\bibitem[\protect\citeauthoryear{Wang, Zheng, Jiang, and Ren}{Wang
  et~al\mbox{.}}{2018b}]%
        {wang2018toward}
\bibfield{author}{\bibinfo{person}{Cong Wang}, \bibinfo{person}{Yifeng Zheng},
  \bibinfo{person}{Jinghua Jiang}, {and} \bibinfo{person}{Kui Ren}.}
  \bibinfo{year}{2018}\natexlab{b}.
\newblock \showarticletitle{Toward privacy-preserving personalized
  recommendation services}.
\newblock \bibinfo{journal}{\emph{Engineering}} \bibinfo{volume}{4},
  \bibinfo{number}{1} (\bibinfo{year}{2018}), \bibinfo{pages}{21--28}.
\newblock


\bibitem[\protect\citeauthoryear{Wang, Zhang, Wang, Zhao, Li, Xie, and
  Guo}{Wang et~al\mbox{.}}{2018a}]%
        {wang2018ripplenet}
\bibfield{author}{\bibinfo{person}{Hongwei Wang}, \bibinfo{person}{Fuzheng
  Zhang}, \bibinfo{person}{Jialin Wang}, \bibinfo{person}{Miao Zhao},
  \bibinfo{person}{Wenjie Li}, \bibinfo{person}{Xing Xie}, {and}
  \bibinfo{person}{Minyi Guo}.} \bibinfo{year}{2018}\natexlab{a}.
\newblock \showarticletitle{Ripplenet: Propagating user preferences on the
  knowledge graph for recommender systems}. In
  \bibinfo{booktitle}{\emph{Proceedings of the 27th ACM International
  Conference on Information and Knowledge Management}}.
  \bibinfo{pages}{417--426}.
\newblock


\bibitem[\protect\citeauthoryear{Wang, Zhang, Wu, Ma, Hong, and Wang}{Wang
  et~al\mbox{.}}{2021}]%
        {wang2021priviledgepgd}
\bibfield{author}{\bibinfo{person}{Shuai Wang}, \bibinfo{person}{Kun Zhang},
  \bibinfo{person}{Le Wu}, \bibinfo{person}{Haiping Ma},
  \bibinfo{person}{Richang Hong}, {and} \bibinfo{person}{Meng Wang}.}
  \bibinfo{year}{2021}\natexlab{}.
\newblock \showarticletitle{Privileged Graph Distillation for Cold Start
  Recommendation}. In \bibinfo{booktitle}{\emph{{SIGIR} '21: The 44th
  International {ACM} {SIGIR} Conference on Research and Development in
  Information Retrieval, Virtual Event, Canada, July 11-15, 2021}},
  \bibfield{editor}{\bibinfo{person}{Fernando Diaz}, \bibinfo{person}{Chirag
  Shah}, \bibinfo{person}{Torsten Suel}, \bibinfo{person}{Pablo Castells},
  \bibinfo{person}{Rosie Jones}, {and} \bibinfo{person}{Tetsuya Sakai}} (Eds.).
  \bibinfo{publisher}{{ACM}}, \bibinfo{pages}{1187--1196}.
\newblock
\urldef\tempurl%
\url{https://doi.org/10.1145/3404835.3462929}
\showDOI{\tempurl}


\bibitem[\protect\citeauthoryear{Wang, He, Cao, Liu, and Chua}{Wang
  et~al\mbox{.}}{2019a}]%
        {wang2019kgat}
\bibfield{author}{\bibinfo{person}{Xiang Wang}, \bibinfo{person}{Xiangnan He},
  \bibinfo{person}{Yixin Cao}, \bibinfo{person}{Meng Liu}, {and}
  \bibinfo{person}{Tat-Seng Chua}.} \bibinfo{year}{2019}\natexlab{a}.
\newblock \showarticletitle{Kgat: Knowledge graph attention network for
  recommendation}. In \bibinfo{booktitle}{\emph{Proceedings of the 25th ACM
  SIGKDD International Conference on Knowledge Discovery \& Data Mining}}.
  \bibinfo{pages}{950--958}.
\newblock


\bibitem[\protect\citeauthoryear{Wang, He, Wang, Feng, and Chua}{Wang
  et~al\mbox{.}}{2019b}]%
        {wang2019neuralngcf}
\bibfield{author}{\bibinfo{person}{Xiang Wang}, \bibinfo{person}{Xiangnan He},
  \bibinfo{person}{Meng Wang}, \bibinfo{person}{Fuli Feng}, {and}
  \bibinfo{person}{Tat-Seng Chua}.} \bibinfo{year}{2019}\natexlab{b}.
\newblock \showarticletitle{Neural graph collaborative filtering}. In
  \bibinfo{booktitle}{\emph{Proceedings of the 42nd international ACM SIGIR
  conference on Research and development in Information Retrieval}}.
  \bibinfo{pages}{165--174}.
\newblock


\bibitem[\protect\citeauthoryear{Wang, Wang, Xu, He, Cao, and Chua}{Wang
  et~al\mbox{.}}{2019c}]%
        {wang2019explainableknowledgegraphkprn}
\bibfield{author}{\bibinfo{person}{Xiang Wang}, \bibinfo{person}{Dingxian
  Wang}, \bibinfo{person}{Canran Xu}, \bibinfo{person}{Xiangnan He},
  \bibinfo{person}{Yixin Cao}, {and} \bibinfo{person}{Tat-Seng Chua}.}
  \bibinfo{year}{2019}\natexlab{c}.
\newblock \showarticletitle{Explainable reasoning over knowledge graphs for
  recommendation}. In \bibinfo{booktitle}{\emph{Proceedings of the AAAI
  Conference on Artificial Intelligence}}, Vol.~\bibinfo{volume}{33}.
  \bibinfo{pages}{5329--5336}.
\newblock


\bibitem[\protect\citeauthoryear{Wu, Wang, Gao, Chen, Fu, Qiu, and He}{Wu
  et~al\mbox{.}}{2022}]%
        {wu2022effectivenessbprbce}
\bibfield{author}{\bibinfo{person}{Jiancan Wu}, \bibinfo{person}{Xiang Wang},
  \bibinfo{person}{Xingyu Gao}, \bibinfo{person}{Jiawei Chen},
  \bibinfo{person}{Hongcheng Fu}, \bibinfo{person}{Tianyu Qiu}, {and}
  \bibinfo{person}{Xiangnan He}.} \bibinfo{year}{2022}\natexlab{}.
\newblock \showarticletitle{On the Effectiveness of Sampled Softmax Loss for
  Item Recommendation}.
\newblock \bibinfo{journal}{\emph{arXiv preprint arXiv:2201.02327}}
  (\bibinfo{year}{2022}).
\newblock


\bibitem[\protect\citeauthoryear{Wu, Zhang, Gao, Yan, and Zha}{Wu
  et~al\mbox{.}}{2021}]%
        {wu2021towardsidcf}
\bibfield{author}{\bibinfo{person}{Qitian Wu}, \bibinfo{person}{Hengrui Zhang},
  \bibinfo{person}{Xiaofeng Gao}, \bibinfo{person}{Junchi Yan}, {and}
  \bibinfo{person}{Hongyuan Zha}.} \bibinfo{year}{2021}\natexlab{}.
\newblock \showarticletitle{Towards Open-World Recommendation: An Inductive
  Model-based Collaborative Filtering Approach}. In
  \bibinfo{booktitle}{\emph{Proceedings of the 38th International Conference on
  Machine Learning, {ICML} 2021, 18-24 July 2021, Virtual Event}}
  \emph{(\bibinfo{series}{Proceedings of Machine Learning Research},
  Vol.~\bibinfo{volume}{139})}, \bibfield{editor}{\bibinfo{person}{Marina
  Meila} {and} \bibinfo{person}{Tong Zhang}} (Eds.).
  \bibinfo{publisher}{{PMLR}}, \bibinfo{pages}{11329--11339}.
\newblock
\urldef\tempurl%
\url{http://proceedings.mlr.press/v139/wu21j.html}
\showURL{%
\tempurl}


\bibitem[\protect\citeauthoryear{Xu, Jin, and Zhou}{Xu et~al\mbox{.}}{2013}]%
        {xu2013speedupimc}
\bibfield{author}{\bibinfo{person}{Miao Xu}, \bibinfo{person}{Rong Jin}, {and}
  \bibinfo{person}{Zhi-Hua Zhou}.} \bibinfo{year}{2013}\natexlab{}.
\newblock \showarticletitle{Speedup matrix completion with side information:
  Application to multi-label learning}. In \bibinfo{booktitle}{\emph{Advances
  in neural information processing systems}}. \bibinfo{pages}{2301--2309}.
\newblock


\bibitem[\protect\citeauthoryear{Yang and Dong}{Yang and Dong}{2020}]%
        {yang2020hagerec}
\bibfield{author}{\bibinfo{person}{Zuoxi Yang} {and} \bibinfo{person}{Shoubin
  Dong}.} \bibinfo{year}{2020}\natexlab{}.
\newblock \showarticletitle{HAGERec: hierarchical attention graph convolutional
  network incorporating knowledge graph for explainable recommendation}.
\newblock \bibinfo{journal}{\emph{Knowledge-Based Systems}}
  \bibinfo{volume}{204} (\bibinfo{year}{2020}), \bibinfo{pages}{106194}.
\newblock


\bibitem[\protect\citeauthoryear{Ying, He, Chen, Eksombatchai, Hamilton, and
  Leskovec}{Ying et~al\mbox{.}}{2018}]%
        {ying2018graphpinsage}
\bibfield{author}{\bibinfo{person}{Rex Ying}, \bibinfo{person}{Ruining He},
  \bibinfo{person}{Kaifeng Chen}, \bibinfo{person}{Pong Eksombatchai},
  \bibinfo{person}{William~L Hamilton}, {and} \bibinfo{person}{Jure Leskovec}.}
  \bibinfo{year}{2018}\natexlab{}.
\newblock \showarticletitle{Graph convolutional neural networks for web-scale
  recommender systems}. In \bibinfo{booktitle}{\emph{Proceedings of the 24th
  ACM SIGKDD International Conference on Knowledge Discovery \& Data Mining}}.
  \bibinfo{pages}{974--983}.
\newblock


\bibitem[\protect\citeauthoryear{Zhang, Chen, Zhang, Xu, and Gao}{Zhang
  et~al\mbox{.}}{2022}]%
        {zhang2022geometricgimc}
\bibfield{author}{\bibinfo{person}{Chengkun Zhang}, \bibinfo{person}{Hongxu
  Chen}, \bibinfo{person}{Sixiao Zhang}, \bibinfo{person}{Guandong Xu}, {and}
  \bibinfo{person}{Junbin Gao}.} \bibinfo{year}{2022}\natexlab{}.
\newblock \showarticletitle{Geometric Inductive Matrix Completion: {A}
  Hyperbolic Approach with Unified Message Passing}. In
  \bibinfo{booktitle}{\emph{{WSDM} '22: The Fifteenth {ACM} International
  Conference on Web Search and Data Mining, Virtual Event / Tempe, AZ, USA,
  February 21 - 25, 2022}}, \bibfield{editor}{\bibinfo{person}{K.~Selcuk
  Candan}, \bibinfo{person}{Huan Liu}, \bibinfo{person}{Leman Akoglu},
  \bibinfo{person}{Xin~Luna Dong}, {and} \bibinfo{person}{Jiliang Tang}}
  (Eds.). \bibinfo{publisher}{{ACM}}, \bibinfo{pages}{1337--1346}.
\newblock
\urldef\tempurl%
\url{https://doi.org/10.1145/3488560.3498402}
\showDOI{\tempurl}


\bibitem[\protect\citeauthoryear{Zhang, Yao, Yu, Huang, Song, Chen, Jiang, and
  Chawla}{Zhang et~al\mbox{.}}{2021}]%
        {zhang2021inductiveicp}
\bibfield{author}{\bibinfo{person}{Chuxu Zhang}, \bibinfo{person}{Huaxiu Yao},
  \bibinfo{person}{Lu Yu}, \bibinfo{person}{Chao Huang},
  \bibinfo{person}{Dongjin Song}, \bibinfo{person}{Haifeng Chen},
  \bibinfo{person}{Meng Jiang}, {and} \bibinfo{person}{Nitesh~V Chawla}.}
  \bibinfo{year}{2021}\natexlab{}.
\newblock \showarticletitle{Inductive Contextual Relation Learning for
  Personalization}.
\newblock \bibinfo{journal}{\emph{ACM Transactions on Information Systems
  (TOIS)}} \bibinfo{volume}{39}, \bibinfo{number}{3} (\bibinfo{year}{2021}),
  \bibinfo{pages}{1--22}.
\newblock


\bibitem[\protect\citeauthoryear{Zhang and Chen}{Zhang and Chen}{2019}]%
        {zhang2019inductiveigmc}
\bibfield{author}{\bibinfo{person}{Muhan Zhang} {and} \bibinfo{person}{Yixin
  Chen}.} \bibinfo{year}{2019}\natexlab{}.
\newblock \showarticletitle{Inductive Matrix Completion Based on Graph Neural
  Networks}. In \bibinfo{booktitle}{\emph{International Conference on Learning
  Representations}}.
\newblock


\end{thebibliography}
\appendix


\section{Aggregators}
\label{app:agg}
In the following we formally define the different aggregators used in the literature that we initially explored in our architecture.
\begin{itemize}[leftmargin=*]
    \item \textit{GCN aggregator}~\cite{kipf2016semigcn} summing the two vectors and applying a nonlinear transformation:
    \begin{equation}
        f_{GCN} = \text{LeakyReLU}(\cmatrix{W}(\hidden{\head} + \hidden{\mathcal{N}_\head}))
    \end{equation}
    \item \textit{GraphSAGE aggregator}~\cite{hamilton2017inductive} concatenating the two vectors before applying a nonlinear transformation:
    \begin{equation}
        f_{GS} = \text{LeakyReLU}(\cmatrix{W}(\hidden{\head} \| \hidden{\mathcal{N}_\head}))
    \end{equation}
    \item \textit{LightGCN aggregator}~\cite{he2020lightgcn} removing all transformations and simply propagating the ego-network vector: 
    \begin{equation}
        f_{LGCN} = \hidden{\mathcal{N}_\head}
    \end{equation}
\end{itemize}

\section{Ablation studies}
\label{app:abl}
We study the effect of different components and parameters in the cold-start setting. 
In particular, we investigate the effect of a gating mechanism, the autoencoder and using only a bipartite graph. 
In both \autoref{table:ablation_ae} and \autoref{table:gate}, bold is the best performing method and `*' signifies statistical significant improvement over the next best performing model. Furthermore we define Cov as 

\begin{equation}
    \text{Cov}@k = \frac{|\bigcup_{\user \in \users} \text{Recommendations}@k(\user)|}{|\recs|}
\end{equation}
where $\text{Recommendations}@k(\user)$ is the set of the top-k items recommended given to a user $\user$. 
The na\"ive TopPop would always have a score of $\frac{k}{|\recs|}$, while a random model would recommend all items given enough users, leading to a coverage score of $1$. 
Intuitively, the metric therefore defines the diversity in a methods recommendations without taking the quality of the recommendations into account.
We note that it is impossible to calculate statistical significance for coverage as all users have $k$ unique items and therefore does not change between methods.

\subsection{Gating mechanism and KG structure}


\para{Gates.}
The results of the method with different gating mechanisms can be seen in \autoref{table:gate}.
In the table, `w/o relation' is the gating mechanism without relation type, i.e., a single type shared for all edges, and `w/o gates' is the method without the gating mechanism.
\begin{itemize}
    \item Overall, the \textit{gating mechanism improves performance}, as we can adaptively select information from neighboring nodes seeing it outperform the two other models in all metrics. 
    \item Disregarding relation types leads to worse performance on all datasets and completely removing the gates leads to dramatically lower performance. 
    \item The models performance without gates is worse than PinSAGE, though still better than IDCF. Only using the users interactions without a gating mechanism, is therefore still better than the reconstruction using in IDCF.
    \item Even without relation types, we see large and statistical significant increase in performance. Allowing the model to select information based on itself and its neighbors are therefore very relevant for the models performance.
\end{itemize}

\para{Effect of the KG.} We create a version of the method where it only uses the collaborative graph instead of the \gls{ckg}, named Bipartite in \autoref{table:gate}.

\begin{itemize}
    \item We observe that the bipartite model performs very well on the \gls{ml} dataset, and less so on the \gls{ab} dataset. 
    The \gls{ab} dataset is less dense making edges added by the \gls{kg} more important when clustering users according to their preferences.
    \item Even with its almost equivalent performance w.r.t. NDCG, the bipartite model lacks diversity. 
    Furthermore, on all datasets, the \gls{own} will outperform the bipartite model on at least one metric with statistical significance.
    On the \gls{ab} we see a large difference in the coverage of the model compared to the bipartite model. 
    This is also visible on the \gls{ml} datasets, though less prevalent. 
    \item Combining the Cov and NDCG metric, we see that using \gls{kg} information leads to more diverse recommendations without degrading the recommendation performance. 
    Having highest diversity while having the highest ranking scores, means \gls{own} is able to recommend diverse and relevant items, with higher degree than without \gls{kg} information.
    The recommendations are therefore less popularity biased when using \gls{kg} information as it recommends a larger set of items.
\end{itemize}

\subsection{Autoencoder}
We study the impact of the \gls{ae} loss, finding it to have minimal effect or decreasing performance (see \autoref{table:ablation_ae}). 
When comparing the best performing model where $\lambda\neq0$ with the model where $\lambda = 0$, we often see no statistical significant increase or decrease.

\begin{itemize}
    \item We initially hypothesized that the encoded features trained without the \gls{ae} loss would overfit and that the model would have a difficult time extracting relevant information through multiple \gls{gnn} layers. 
    The \gls{ae} loss would therefore help with extracting information while allowing the recommendation loss to tailor the encoded embedding to the recommendation setting.
    Yet, the results does not indicate that the \gls{ae} loss improves the performance for our model.
    \item On the \gls{ml} and \gls{ab} datasets, we see a statistic significant decrease in performance between $\lambda = 0$ and $\lambda = 2$. 
    A high \gls{ae} loss therefore does not give better recommendation capabilities. 
    \item Lower values of $\lambda$ either decreases performance or maintains it without statistical significant increases. 
    We therefore leave it as future work to find out if it is possible to utilize the \gls{ae} loss in a meaningful way for similar methods.
\end{itemize}

\begin{table*}[!htb]
\centering
\caption{Effect of autoencoder loss measured at 20.}
\label{table:ablation_ae}
\vspace*{-12pt}
\resizebox{1\textwidth}{!}{%
\begin{tabular}{l|rrrr|rrrr|rrrr}
  
 & \multicolumn{4}{c|}{\textbf{MovieLens Subsampled + 1250 users}} & \multicolumn{4}{c|}{\textbf{MovieLens Subsampled + 90\%}} & \multicolumn{4}{c}{\textbf{Amazon Book Subsampled + 15\%}} \\ \hline
 
 \multicolumn{1}{c|}{\textbf{$\lambda$}}
 & \multicolumn{1}{c}{\textbf{NDCG}} & \multicolumn{1}{c}{\textbf{Recall}} & \multicolumn{1}{c}{\textbf{Precision}} & \multicolumn{1}{c|}{\textbf{Cov}} 
 & \multicolumn{1}{c}{\textbf{NDCG}} & \multicolumn{1}{c}{\textbf{Recall}} & \multicolumn{1}{c}{\textbf{Precision}} & \multicolumn{1}{c|}{\textbf{Cov}} 
 & \multicolumn{1}{c}{\textbf{NDCG}} & \multicolumn{1}{c}{\textbf{Recall}} & \multicolumn{1}{c}{\textbf{Precision}} & \multicolumn{1}{c}{\textbf{Cov}} \\ \hline
\textbf{0.00} 
    & \textbf{0.18123} & 0.18123 & \textbf{0.08292} & 0.19831 
    & \textbf{0.18367}* & \textbf{0.24374}* & \textbf{0.08335} & \textbf{0.44127}
    & 0.06842 & \textbf{0.14469} & \textbf{0.01171} & 0.20390\\ \hline
\textbf{0.01} 
    & 0.18013 & \textbf{0.23694} & 0.08232 & \textbf{0.20402} 
    & 0.18260 & 0.24241 & 0.08320 & 0.43915 
    & 0.06745 & 0.14217 & 0.01141 & 0.17612\\ \hline
\textbf{0.10} 
    & 0.17642 & 0.23271 & 0.08152 & 0.17989 
    & 0.17947 & 0.23806 & 0.08138 & 0.37545 
    & 0.06812 & 0.14230 & 0.01166 & \textbf{0.24278}\\ \hline
\textbf{0.50} 
    & 0.17708 & 0.23583 & 0.08076 & 0.17016 
    & 0.17818 & 0.23729 & 0.08139 & 0.33672 
    & \textbf{0.06861} & 0.14270 & 0.01170 & 0.22322\\ \hline
\textbf{1.00} 
    & 0.16832 & 0.22822 & 0.07860 & 0.15302 
    & 0.17339 & 0.23201 & 0.07925 & 0.29397 
    & 0.06251 & 0.13347 & 0.01089 & 0.18574\\ \hline
\textbf{2.00} 
    & 0.17089 & 0.22536 & 0.07736 & 0.14857 
    & 0.17295 & 0.23021 & 0.07838 & 0.27429 
    & 0.05939 & 0.12721 & 0.01034 & 0.13888\\ \hline
\end{tabular}%
}
\end{table*}

\begin{table*}[!htb]
\centering
\caption{Effect of the gating mechanism and KG measured at 20.}
\label{table:gate}
\vspace*{-12pt}
\resizebox{1\textwidth}{!}{%
\begin{tabular}{l|rrrr|rrrr|rrrr}
 & \multicolumn{4}{c|}{\textbf{MovieLens Subsampled + 1250 users}} & \multicolumn{4}{c|}{\textbf{MovieLens Subsampled + 90\%}} & \multicolumn{4}{c}{\textbf{Amazon Book Subsampled + 15\%}} \\ \hline
 & \multicolumn{1}{c}{\textbf{NDCG}} & \multicolumn{1}{c}{\textbf{Recall}} & \multicolumn{1}{c}{\textbf{Precision}} & \multicolumn{1}{c|}{\textbf{Cov}} 
 & \multicolumn{1}{c}{\textbf{NDCG}} & \multicolumn{1}{c}{\textbf{Recall}} & \multicolumn{1}{c}{\textbf{Precision}} & \multicolumn{1}{c|}{\textbf{Cov}} 
 & \multicolumn{1}{c}{\textbf{NDCG}} & \multicolumn{1}{c}{\textbf{Recall}} & \multicolumn{1}{c}{\textbf{Precision}} & \multicolumn{1}{c}{\textbf{Cov}} \\ \hline
\textbf{Bipartite} 
    & \textbf{0.18331} & 0.24071 & 0.08264 & 0.17778 
    & \textbf{0.18379} & \textbf{0.24341} & \textbf{0.08309} & \textbf{0.42455} 
    & 0.06386 & \textbf{0.13484} & \textbf{0.01108} & \textbf{0.18751}\\ \hline
\textbf{W/o relations} 
    & 0.17720 & \textbf{0.24193} & \textbf{0.08288} & \textbf{0.18243} 
    & 0.17799 & 0.23688 & 0.08140 & 0.34519
    & \textbf{0.06451} & 0.13395 & 0.01106 & 0.18465\\ \hline
\textbf{W/o gates} 
    & 0.12482 & 0.17000 & 0.05848 & 0.08148 
    & 0.09416 & 0.13479 & 0.04331 & 0.05122 
    & 0.03653 & 0.00634 & 0.00927 & 0.07673  \\ \hline\hline
\textbf{SimpleRec} 
    & 0.18085 & \textbf{0.24618} & \textbf{0.08460}* & \textbf{0.18540} 
    & \textbf{0.18411} & \textbf{0.24482}* & \textbf{0.08365}* & \textbf{0.42646} 
    & \textbf{0.06472} & \textbf{0.14055}* & \textbf{0.01144}* & \textbf{0.21460}  \\ \hline
\end{tabular}%
}
\vspace*{-12pt}
\end{table*}



\end{document}